\documentclass{article}

\usepackage[
letterpaper, 
total = {7in}
]{geometry}

\usepackage{bm}

\usepackage{amsfonts}

\usepackage{amsmath}

\usepackage{amssymb}

\usepackage{fancyhdr}

\usepackage{parskip}

\usepackage[T1]{fontenc}

\usepackage{dcolumn}

\usepackage{multirow}

\usepackage{comment}

\usepackage{caption}

\usepackage{subcaption}

\usepackage{appendix}

\usepackage{xcolor}

\usepackage{graphicx}

\usepackage{multirow}

\usepackage{tikz}

\usepackage{hyperref}

\usepackage{url}

\usepackage{esint}

\usepackage{tensor}

\usepackage{xspace}

\usepackage{authblk}

\usepackage{listings}

\definecolor{codegray}{rgb}{0.5,0.5,0.5}
\definecolor{codered}{rgb}{0.75,0,0}
\definecolor{backcolour}{rgb}{0.95,0.95,0.92}
\lstdefinestyle{mystyle}{
    backgroundcolor = \color{backcolour},
    keywordstyle = \color{blue},
    numberstyle = \tiny\color{codegray},
    stringstyle = \color{codered},
    basicstyle = \ttfamily\footnotesize,
    breakatwhitespace = false,         
    captionpos = b,                    
    keepspaces = true,                 
    numbers = left,                    
    numbersep = 5pt,                  
    showspaces = false,                
    showstringspaces = false,
    showtabs = false,                  
    tabsize = 2
}
\newcommand{\italics}[1]{\textit{#1}}
\newcommand{\boldface}[1]{\textbf{#1}}
\newcommand{\quotesaround}[1]{``#1''}

\newcommand{\parentheses}[1]{\left( #1 \right)}
\newcommand{\squarebrackets}[1]{\left[ #1 \right]}
\newcommand{\curlybrackets}[1]{\left\{ #1 \right\}}

\newcommand{\absoluteValue}[1]{\left|#1\right|}

\newcommand{\vectorarrow}[1]{\vec{#1}}

\newcommand{\dotproduct}[2]{#1 \cdot #2}

\title{A Canonical Lagrangian Formulation of the Two-Dimensional Lotka-Volterra System}

\author{Dima Watkins
\thanks{e-mail: \href{mailto:bem8mq@virginia.edu}{bem8mq@virginia.edu}
}}
\author{Gene Chen
\thanks{e-mail: \href{mailto:wkp5pp@virginia.edu}{wkp5pp@virginia.edu}
}}

\affil[1]{Department of Physics, University of Virginia, Charlottesville, Virginia 22904-4714, USA}

\date{\today}

\begin{document}

\maketitle

\begin{abstract}  

Hamiltonian and Lagrangian mechanics are powerful frameworks for analyzing physical systems. Previous work has extended these formalisms to ecological systems, such as the predator-prey Lotka-Volterra (LV) system. In this Article, we derive a canonical Lagrangian for the two-dimensional LV model directly from its Hamiltonian representation. We find that the two-dimensional LV system admits a standard canonical Lagrangian formulation with one degree of freedom and a non-quadratic kinetic structure. This formulation admits a mechanical interpretation in terms of a particle moving in a potential well, where the non-standard kinetic structure produces a position-dependent damping term that can instead act as \quotesaround{revving.} The derivation provides a direct connection between predator-prey dynamics and a canonical formulation of mechanical dynamics. As a verification of the construction, we apply Noether's procedure to the explicitly time-independent derived Lagrangian and reveal that the well-known Hamiltonian of the LV system is the corresponding conserved quantity. We also uncover a subtle redundancy associated with the choice of canonical momentum and its identification with the original population variables.

\end{abstract}


\section{Introduction}
\label{sec:introduction}

What we now call the Lotka-Volterra system of equations was first introduced by Alfred J. Lotka in the modeling of autocatalytic chemical reactions \cite{lotka-autocatalysis-equations}. Vito Volterra independently proposed the same equations to explain why prey fish thrived in the Adriatic Sea during the First World War, a period when conflict had slowed commercial fishing \cite{volterra-population-competition, summary-of-nonlinear-population-models}. We now refer to the two-dimensional dynamical system of equations as the \quotesaround{Lotka-Volterra Model} \cite{lotka-volterra-unified-model}. This two-dimensional Lotka-Volterra model has seen application in chemistry, mathematical biology and ecology, and recently in economic and marketing theory \cite{economic-lv-application, goodwin-lv-economics, hung-2017-dram-market-lv-model}. It is clear that analysis of its underlying structure not only offers a better understanding of its abstract mathematical form but clearly also enables a variety of rich interpretations. 

In the following Article, we propose a Lagrangian for the two-dimensional Lotka-Volterra system (hereafter LV system). It is interesting that a Lagrangian for this system can be constructed, as the LV system is not typically thought to describe a standard physical or mechanical system. We shall soon see that this objective --- writing a Lagrangian for the two-dimensional LV system --- has been attempted several times before; our approach deviates from precedent. While the low dimensionality of the model in consideration renders this work somewhat elementary, the analysis we offer in this Article is new.

This work was based on the observation by E. Kerner \cite{kerner-lv-hamiltonian-transformation} and recapitulated in 1990 by Y. Nutku \cite{NUTKU199027} (to which E. Kerner responded in \cite{kerner-1990-lv-hamiltonian-stat-mech}) that the Lotka-Volterra system admits a Hamiltonian structure. After Nutku's publication, the $n$-dimensional LV model received tremendous coverage, the $n$ here referring to a system of $n$ interacting populations. (The $n$-dimensional LV model was not unheard of before Nutku: preceding work on the $n$-dimensional model explored its connection with nonlinear lattice models, like the Toda lattice \cite{itoh-1987-integrals-lv-odd}. However, no reference to the \quotesaround{Hamiltonianization} of the model was made in Ref. \cite{itoh-1987-integrals-lv-odd}, in contrast to Refs. \cite{kerner-lv-hamiltonian-transformation, NUTKU199027} and others.) Much of the \quotesaround{Hamiltonianization} research program seeks to find first integrals for the $n$-dimensional generalization, and will often refer to these as Hamiltonian(s); M. Plank explored the general question of the Hamiltonian structure of the $n$-dimensional LV equations \cite{plank-1995-n-dim-lv-hamiltonian}, which received responses from E. Kerner in Ref. \cite{kerner-1997-comment-n-dim-lv} and L. Cairó and M. R. Feix in Ref. \cite{cairo-1996-comments-n-dim-lv-hamiltonian}, reminding the field about extant literature on the $n$-dimensional system; M. Plank continued in Ref. \cite{plank-1996-bihamiltonian-lv-3d} and identified a bi-Hamiltonian structure (see Definition 2 therein) for the $n = 3$ LV model; Ref \cite{ballesteros-2011-deformations-lv} further explores the $n = 3$ LV system and uncovers its Poisson-Lie Group structure; Ref. \cite{christodoulides-2009-darboux-lv-3d} studied Darboux polynomials for the $3$-dimensional LV system; Ref. \cite{bermejo-1998-hamiltonian-darboux} casts the $n$-dimensional problem into one with the structure of a Poisson manifold that allows one to essentially analyze Hamiltonian-like systems on manifolds of odd dimensionality, something that a symplectic manifold does not allow for: in their words, \quotesaround{\dots the
classical restriction to even-dimensional manifolds is not present in Poisson systems} (\cite{bermejo-1998-hamiltonian-darboux} pg. 6163).

It is evident that the $n$-dimensional generalization of the Lotka-Volterra model is rich in mathematical detail and is an excellent playground to study mathematical concepts and techniques. On the other hand, many of these notions are unheard of in traditional physics formalisms --- such as bi-Hamiltonians. What distinguishes this present work from this paradigm is that it seeks to maintain the formalism of classical mechanics throughout. That is, for example, we hold that phase space shall always be even-dimensional. This proposition is in contrast to Ref. \cite{plank-1995-n-dim-lv-hamiltonian}, pg. 3521, in which, for example, the author exploring $n$-dimensional LV systems remarks that the classical notion of Hamiltonian systems possess an \quotesaround{\dots artificial restriction to even-dimensional phase spaces.}

With the \quotesaround{Hamiltonianization} program of the $n$-dimensional LV system well-established, we wondered if the \quotesaround{Lagrangianization} of the system received similar attention; Given that the LV equations possess a Hamiltonian structure, one may ask if it came from a corresponding Lagrangian via the standard Legendre transform. It was shown in Refs. \cite{kerner-1972-gibbs, nunez-1998-lv-lagrangian, pham-2022-lagrangian-biology} that one can write Lagrangians for $n$-dimensional LV systems, but the proposed Lagrangians in these Articles were not related to Hamiltonians, and the results therein are different in form from what we will derive later.

Our plan for this Article is the following: In Section \ref{sec:lv-system-review}, we speak briefly about the canonical form of the Lotka-Volterra equations; in Section \ref{sec:hamiltonian-structure}, we review the well-established Hamiltonian of the Lotka-Volterra system; in Section \ref{sec:prevous-lagrangian-work}, we discuss previous work on Lagrangians describing the Lotka-Volterra system before we derive ours in Section \ref{sec:our-lagrangian}; with our form of the Lagrangian, we then analyze its mechanics in Section \ref{sec:mechanics-of-lv}. Lastly, in Section \ref{sec:discussion}, we discuss the findings, with an emphasis on a subtlety that emerged from the previous section, and offer a method that sheds light on this issue.

\section{Review of LV System}
\label{sec:lv-system-review}

(As remarked in Section \ref{sec:introduction}, the LV model has found application in many scientific disciplines. Henceforward, we shall describe the model in its ecological context, which will be clarified through our notation and terminology shortly.) The most general form of the two-dimensional LV equations governing the dynamics of the area-densities of a species of prey $x$ and predators $y$ is written below, and all the model parameters involved are positive:

\begin{equation}
\label{eq:lotka-volterra-equations}
    \frac{dx}{dt} = \alpha x - \beta x y, \quad
    \frac{dy}{dt} = - \gamma y + \delta x y.
\end{equation}

(It has been remarked in Ref. \cite{leconte-2022-limit-cycles-lotka-voltera-plasma} and is an exercise in Ref. \cite{strogatz-2018-nonlinear-dynamics} that three of the four model parameters can be eliminated by appropriate rescaling of $x$, $y$, and $t$. In this analysis, we proceed with all four parameters. However, for simplicity, we can bring the units of $x$ and $y$ from population per area to having no dimension by imposing that we work with a fixed area.) As can be seen, without predator-prey interactions, the prey population exponentially increases with a characteristic growth scale $\alpha$ and the predator population exponentially decreases with $\gamma$. Without interactions, the exponential growth and decay continue without bound, leading to the extinction of predators ($y \to 0$) and an infinite number of  prey ($x \to \infty$). To introduce inter-species interactions, the per-capita growth rate of each population ($\dot{x}_{i}/x$, where $\dot{x}_{i} = dx_{i}/dt$, and $x_{i}$ is either $x$ or $y$ in Eqs. \ref{eq:lotka-volterra-equations}) is assumed to follow modified logistic growth: each population's intrinsic growth rate is a function of the total number of the competing species; this procedure results in the non-linear, non-reciprocal inter-species coupling terms, $\beta x y$ and $\delta x y$, and results in the two-dimensional Lotka-Volterra equations.

\section{Hamiltonian Structure}
\label{sec:hamiltonian-structure}

We now discuss how the Hamiltonian structure emerges from the original LV equations. The first step in this construction is to identify a constant of the motion in Eqs. \ref{eq:lotka-volterra-equations}; even Lotka was aware of this constant of the motion when he came up with the model \cite{evans-1999-new-transformation-lv, lotka-1920-undamped-oscillations-mass-action}. This dynamical invariant called the \quotesaround{Hamiltonian} of the system in Refs. \cite{NUTKU199027, baumann-1991-symmetries-lv-model}:

\begin{equation}
\label{eq:lv-system-const-of-motion}
    C = \delta x - \gamma \ln \parentheses{x} + \beta y - \alpha \ln \parentheses{y}.
\end{equation}

We have denoted the conserved quantity with $C$ instead of $H$ because one cannot recover the dynamics described by Eqs. \ref{eq:lotka-volterra-equations} by naïvely applying (what is thought to represent) Hamilton's Equations with respect to Eq. \ref{eq:lv-system-const-of-motion}. That is,

\begin{equation}
    \frac{\partial C}{\partial x} = \delta - \frac{\gamma}{x} \neq \dot{y}, \quad
    \frac{\partial C}{\partial y} = \beta - \frac{\alpha}{y} \neq -\dot{x},
\end{equation}

where the equations for $\dot{x}$ and $\dot{y}$ are, of course, those pertaining to the original system. (Above, when we applied \quotesaround{Hamilton's Equations,} we assumed that $x$ played the role of $p$ and $y$ played the role of $q$, as evidenced by the relative minus sign in front of $\dot{x}$. However, there is no a priori reason to make this assumption. This freedom will be discussed in sections to come.) Equivalently, Eqs. \ref{eq:lotka-volterra-equations} cannot be generated through naïve application of the Poisson Brackets with respect to $C$, i.e. $\curlybrackets{x, C} \neq \dot{x}$ and the same for $y$. On the other hand, from these considerations, one can glean how to modify the bracket structure to correctly derive the original equations; it involves the following definition:

\begin{equation}
\label{eq:modified_lk_possion_bracket}
    \curlybrackets{f, g}^{'}\parentheses{x, y} := - xy \parentheses{ \frac{\partial f}{\partial x} \frac{\partial g}{\partial y} - \frac{\partial g}{\partial x} \frac{\partial f}{\partial y} },
\end{equation}

where the prime $\parentheses{'}$ indicates that these Poisson Brackets are exotic in some sense; that is, computation of $\curlybrackets{x, C}'$ and $\curlybrackets{y, C}'$ will now produce Eqs. \ref{eq:lotka-volterra-equations}. That we cannot recover the LV equations with the standard Hamiltonian formalism suggests that we have not chosen canonical coordinates.

Previous literature (Refs. \cite{kerner-1990-lv-hamiltonian-stat-mech, fernandes-1998-lotka-volterra-hamiltonian, baigent-2010-lotka, cassamchenai-2024-simple-conditions}) has found that there exists a transformation from the current set of coordinates ($x$ and $y$) to a set of canonical coordinates ($p$ and $q$), rectifying the above pathological Poisson Bracket (Eq. \ref{eq:modified_lk_possion_bracket}). Technically, in the work of E. H. Kerner (Refs. \cite{kerner-lv-hamiltonian-transformation, kerner-1990-lv-hamiltonian-stat-mech, kerner-1972-gibbs}), the logarithmic transformation also involved normalizing the dynamical variables by their stationary values. That is, there is precedent in performing the following coordinate transformation:

\begin{align}
    p &:= \ln \parentheses{x} \label{eq:lv-x-to-p-transformation} \\
    q &:= \ln \parentheses{y} \label{eq:lv-y-to-q-transformation}
\end{align}

Note that, due to the symmetry of the transformation, the association of $x$ with $p$ and $y$ with $q$ is largely arbitrary; the opposite assignment would also work provided we also modify the pathological Poisson Bracket accordingly (i.e. if the overall multiplicative factor in Eq. \ref{eq:modified_lk_possion_bracket} is $-xy$ or $xy$.) Another implication of choosing a different association between the pairs $\parentheses{x,y}$ and $\parentheses{q, p}$ is that their physical  interpretation would change accordingly, which we shall now comment on. In the LV system, it is understood that $x$ and $y$ only have a physical interpretation if they are greater than zero; they represent the, say, rabbit and wolf population densities respectively. Thus, the transformation above is, in this sense, invertible. Additionally, since the logarithmic function maps non-negative real numbers to the entire real line, this mapping thus is natural: in Hamiltonian mechanics, $p$ and $q$ typically range across the whole real line (exceptions include, e.g., topological identification of generalized coordinates, such as $\theta$ or $\phi$ when describing motion confined on a cylinder or sphere).

Effecting the coordinate transformation above allows us to rewrite the dynamical invariant $C$ as 

\begin{equation}
    \label{eq:hamiltonian-of-lv}
    H_{\text{LV}} = \delta e^{p} - \gamma p + \beta e^{q} - \alpha q,
\end{equation}

where we now call the constant $H$ instead of $C$ (because it generates the LV dynamics according to the canonical Poisson Brackets). Now, by applying Hamilton's Equations either directly or using the canonical form of the Poisson Bracket (which are equivalent), one can generate the \quotesaround{equations of motion} for the LV system. Using the canonical form of Hamilton's Equations yields

\begin{align}
    \dot{q} &= \frac{\partial H_{\text{LV}}}{\partial p} = \delta e^{p} - \gamma \label{eq:lv-hamiltonian-q-dot} \\
    - \dot{p} &= \frac{\partial H_{\text{LV}}}{\partial q} = \beta e^{q} - \alpha  \label{eq:lv-hamiltonian-p-dot},
\end{align}

and these equations are recovered (as they should be) if one computes $\curlybrackets{p, H_{\text{LV}}}$ and $\curlybrackets{q, H_{\text{LV}}}$ using the canonical form of the Poisson Brackets:

\begin{align}
\label{eq:standard-poisson-brackets}
    \curlybrackets{f, g}\parentheses{p, q} = \frac{\partial f}{\partial p} \frac{\partial g}{\partial q} - \frac{\partial g}{\partial p} \frac{\partial f}{\partial q}.
\end{align}

\section{Lagrangian Structure}
\label{sec:lagrangian-structure}

That there exists a Hamiltonian structure to the LV equations, one may ask if there is a corresponding Lagrangian, as we asked in Section \ref{sec:introduction}. We shall construct one in this article according to the rules of Lagrangian and Hamiltonian mechanics. It was first written around 50 years ago that LV-type population dynamics may be spoken of in a Hamiltonian framework (Ref. \cite{kerner-1972-gibbs, kerner-1990-lv-hamiltonian-stat-mech}), but we are aware of only a handful of articles that examined a Lagrangian structure of the system, some explicitly constructing it and others referring to the possibility of constructing it (Refs. \cite{kerner-1972-gibbs, nunez-1998-lv-lagrangian, pham-2022-lagrangian-biology, paine-1982-lagrangian-biomodels, kerner-1959-lv-lagrangian-euler-lagrange}). The previous approaches employed a construction of the Lagrangian that we now briefly recapitulate --- we will not use these methods in this Article.

\subsection{Previous Work}
\label{sec:prevous-lagrangian-work}

Before further discussing the attempt at finding the Lagrangian for the LV system, we will describe generally the problem that is referred to as the \quotesaround{Inverse Problem of Lagrangian Mechanics.} Firstly, one should take care to note that the current Article addresses the following general issue: we are given an arbitrary set of dynamical equations, and we are trying to determine if the equations can be derived from a Hamiltonian and, now, a Lagrangian. In Section \ref{sec:hamiltonian-structure}, we first discussed finding a dynamical invariant in the two-dimensional system, and then we promoted the structure into a Hamiltonian system in canonical coordinates. In this section, we turn to the task of finding a Lagrangian \italics{given} success in writing a Hamiltonian for the LV system (Eq. \ref{eq:hamiltonian-of-lv}). This question is, of course, a natural one: Given the Hamiltonian, can you find the Lagrangian?

We will now discuss some alternative methods to find a Lagrangian for a given dynamical system. The first method involves approaching an arbitrary system of $n$ \italics{second-order} ODEs. This approach is a natural one because the Euler-Lagrange equations produce second-order equations of motion. (Of course, note that the LV equations are first-order ODEs.) Generally, the second-order ODE system is written according to

\begin{equation}
    \ddot{x}_{i} = f_{i} \parentheses{x, \dot{x}; t},
\end{equation}

where $1 \leq i \leq n$ and $\ddot{x} = d\dot{x}/dt$. We now ask, Given that this ODE system has second-order time derivatives, might it have been derived from a Lagrangian using the Principle of Least Action? That is, does there exist an $L\parentheses{x, \dot{x}; t}$ such that the ODE system above are given by the Euler–Lagrange equations?

A collection of extant literature has covered the inverse problem of Lagrangian mechanics from this perspective. Historically, it appears that conditions on the solution to the inverse problem were first spoken of in 1887 by Helmholtz in \cite{helmholtz-1887-conditions} and were further explored by Douglas, J. in 1941 \cite{douglas-1941-inverse-problem}. (The three Helmholtz conditions governing the existence of Lagrangian for an arbitrary second-order ODE system involve the determination of the components of a $n\times n$ nonsingular anti-symmetric matrix. See Ref. \cite{rawashdeh-2006-nilradical-lie-algebra} Sec. IV and Ref. \cite{zenkov-2015-inverse} Ch. 1 for a quick review, and Ref. \cite{saunders-2010-inverse-problem-review} for a 30-year review of the inverse problem. For more thorough treatments of the inverse problem of Lagrangian mechanics, please consult Ref. \cite{anderson-1992-inverse-problem}, and Santilli's textbook \cite{santilli-1978-inverse-problem-newtonian}, which provided a comprehensive work on the problem.) Contemporary literature combines Lie group analysis with PDE theory to examine the symmetries of systems of ODEs to derive corresponding \quotesaround{Lagrangians}; see Refs. \cite{nucci-2008-jacobi-multiplier-lagrangian-oscillator, nucci-2015-symmetries-easter-island, musielak-2020-lagrangian-special-functions} for examples on how these techniques are applied.

In the case where the dynamical system of interest is an arbitrary one of $n$ \italics{first-order} ODEs, one can seek Lagrangians for the system in a slightly different fashion, the main difference being that the Euler-Lagrange equations must now generate first-order ODEs. This method has found application in Refs. \cite{kerner-1972-gibbs, nunez-1998-lv-lagrangian, paine-1982-lagrangian-biomodels, kerner-1959-lv-lagrangian-euler-lagrange, santilli-1978-inverse-problem-newtonian}, and is referred to in Refs. \cite{lumsden-1979-constraint-biosystems}. Typically, the arbitrary dynamical equations are prescribed according to the following equation:

\begin{equation}
    \label{eq:}
    \dot{x}_{i} = X_{i}\parentheses{x;t},
\end{equation}

which is the notation used in Ref. \cite{kerner-1972-gibbs}, pg. 334.


In this Article, we do not appeal to the machinery developed in solving the Inverse Problem of Lagrangian Mechanics. In contrast, we offer the approach defined by first finding a Hamiltonian for the LV system --- which is something the field has known for some time --- and then using the standard Legendre transform to back-calculate the Lagrangian. It appears that this approach has not been attempted yet before (at least, the results we will show in Section \ref{sec:our-lagrangian} have not appeared in the literature to date). To become acquainted with some of the literature on the problem of the LV Lagrangian, we will briefly recapitulate some of the previous results on Lagrangians for the LV system as derived using the machinery from the Inverse Problem of Lagrangian Mechanics.

For the LV equations written according to 

\begin{equation}
    \dot{w}_{1} = w_{1}\parentheses{a + b w_{2}}, \quad \dot{w}_{2} = w_{2}\parentheses{A + B w_{1}},
\end{equation}

Ref. \cite{pham-2022-lagrangian-biology} finds a Lagrangian that generates \italics{second order} ODEs equivalent to the system above, in accordance with the standard Inverse Problem of Lagrangian Mechanics, to be

\begin{equation}
    L_{\text{Pham}} = \frac{1}{2} \parentheses{\frac{\dot{w}_{1}}{w_{1}}}^{2} + \frac{1}{2} \parentheses{\frac{\dot{w}_{2}}{w_{2}}}^{2} - a \parentheses{B w_{1} + A \ln \absoluteValue{w_{1}}} - A \parentheses{b w_{2} + a \ln \absoluteValue{w_{2}}},
\end{equation}

which is just the sum of Ref. \cite{pham-2022-lagrangian-biology}'s Eqs. (19a) and (19b); for the LV equations written according to 

\begin{equation}
    \dot{x} = x\parentheses{a - y}, \quad \dot{y} = -y\parentheses{b - x},
\end{equation}

in which our $\beta = \delta = 1$ (defining reciprocal inter-species coupling), J. Fernández-Núñez (Ref. \cite{nunez-1998-lv-lagrangian}) derived that the Lagrangian generating its dynamics is

\begin{equation}
    L_{\text{Nunez}} = \frac{1}{2} \frac{\ln \absoluteValue{y}}{x} \dot{x} - \frac{1}{2} \frac{\ln \absoluteValue{x}}{y} \dot{y} - \parentheses{a \ln \absoluteValue{y} + b \ln \absoluteValue{x} - x - y},
\end{equation}

which is a transcription of Eq. (5) in Ref. \cite{nunez-1998-lv-lagrangian}; for the LV dynamics written according to 

\begin{equation}
    \dot{x}_{1} = x_{1} - x_{1} x_{2}, \quad \dot{x}_{2} = - x_{2} + x_{1} x_{2},
\end{equation}

in which $\alpha = \beta = \gamma = \delta = 1$ (and for a multiplicative, time-independent, $2 \times 2$, nonsingular anti-symmetric matrix function multiplying the equations above, involved in the approach to solving the inverse problem of the second type), it was found that the Lagrangian that produces the dynamics through the Euler-Lagrange equations is

\begin{equation}
    L_{\text{Paine}} = e^{-\parentheses{x_{1} + x_{2}}}\curlybrackets{\frac{x_{1} \dot{x}_{2} - x_{2} \dot{x}_{1}}{x_{1} + x_{2}} + \frac{x_{1} \dot{x}_{2} - x_{2} \dot{x}_{1}}{\parentheses{x_{1} + x_{2}}^{2}} - x_{1}x_{2}} + \frac{x_{2} \dot{x}_{1} - x_{1} \dot{x}_{2}}{\parentheses{x_{1} + x_{2}}^{2}},
\end{equation}

which is a transcription of Eq. (38) in Ref. \cite{paine-1982-lagrangian-biomodels}; in Ref. \cite{kerner-1972-gibbs} and the embedded article Ref. \cite{kerner-lv-hamiltonian-transformation}, while a Lagrangian for an $n$-dimensional LV system is not explicitly written down, E. H. Kerner provides a general derivation for a Lagrangian for the inverse problem in the latter context; it too is linear in generalized velocities.

\subsection{Our Proposed Lagrangian}
\label{sec:our-lagrangian}

In passing from Hamiltonian systems to Lagrangian systems and back, one's conception of the dynamical system must change as well. The original population dynamics equations describe the change in predator ($y$) and prey ($x$) populations. The Hamiltonian structure (Eq. \ref{eq:hamiltonian-of-lv}) respects this interpretation, but, in using the Hamiltonian formalism, $p$ and $q$ are now to be regarded as conjugate quantities; back-computing a Lagrangian entails the understanding that the interplay of these two dynamical quantities is actually an equivalent description of a single one whose equations of motion are second-order in time. It is here that our work departs from that presented in Section \ref{sec:prevous-lagrangian-work}: the Lagrangians describing the LV dynamics are linear in generalized velocities for both species, and thereby describe two degrees of freedom. (In addition, as observed in Ref. \cite{paine-1982-lagrangian-biomodels}, some of these Lagrangians are singular, the definition of which is provided in Eq. \ref{eq:hessian-definition}.) In the present article, we begin with a Hamiltonian expressing two conjugate quantities, $p$ and $q$, and write a Lagrangian in terms of $q$ and $\dot{q}$ only. In other words, we shall find only one degree of freedom in our approach.

Without suppressing the functional dependencies --- as it is important for our discussion, albeit we will only use it once --- the Hamiltonian, as generated from a given Lagrangian, is defined through a Legendre transform:

\begin{equation}
\label{eq:hamiltonian-from-lagrangian}
    H \parentheses{\vec{p}, \vec{q}; t} = \sum_{i = 1}^{N} p_{i} \dot{q}^{i}\parentheses{\vec{p}, \vec{q}; t} - L \parentheses{\vec{q}\parentheses{\vec{p}, \vec{q}; t}, \dot{\vec{q}}\parentheses{\vec{p}, \vec{q}; t}; t},
\end{equation}

where $N$ is the number of degrees of freedom in the system.

As Lagrangians speak of motion in terms of generalized coordinates and their velocities, Hamiltonians speak of the interplay of generalized coordinates and momenta. It is our task to find the Lagrangian and, in doing so, identify the canonical momentum from which the Hamiltonian was derived. We can, of course, invert Eq.  \ref{eq:hamiltonian-from-lagrangian} to write the Lagrangian of the system using the Hamiltonian. Again, without suppressing the functional structure of the Lagrangian, we seek to generate the LV Lagrangian according to the following equation:

\begin{equation}
\label{eq:lagrangian-from-hamiltonian}
    L \parentheses{\vec{q}, \dot{\vec{q}}; t} = \sum_{i = 1}^{N} p_{i}\parentheses{\vec{q}, \dot{\vec{q}}; t} \dot{q}^{i} - H \parentheses{\vec{q}\parentheses{\vec{q}, \dot{\vec{q}}; t}, \vec{p}\parentheses{\vec{q}, \dot{\vec{q}}; t}; t}.
\end{equation}

Speaking once more of the task at hand, it may now be seen from the equation above that what we must find is $p\parentheses{\vec{q}, \dot{\vec{q}}; t}$, since we already have the Hamiltonian; this remaining piece will enable us to write down the Lagrangian.

In computing the equations of motion from the LV Hamiltonian (Eq. \ref{eq:hamiltonian-of-lv}), we found that Eq. \ref{eq:lv-hamiltonian-q-dot} provides the recipe for identifying the generalized velocity:

\begin{equation}
\label{eq:lv_conjugate_momentum}
    p\parentheses{\dot{q}} = \ln \parentheses{\frac{\dot{q} + \gamma}{\delta}}.
\end{equation}

Note that from Eq. \ref{eq:lv-hamiltonian-q-dot}, we can also guarantee that the argument of the logarithm will always be positive: $\dot{q} + \gamma = \delta e^{p} > 0$, as the exponential will always be positive. This ensures that Eq. \ref{eq:lv_conjugate_momentum} is invertible. We will report the LV Lagrangian momentarily, but in the meanwhile, it is worth mentioning that another check for the condition of invertibility is the nonvanishing of the determinant of the Hessian associated with generalized velocities in the Lagrangian --- the Hessian condition \cite{jose-1998-classical-dynamics}. That is,

\begin{equation}
\label{eq:hessian-definition}
    \det \absoluteValue{\frac{\partial^{2} L}{\partial \dot{q}^{j} \partial \dot{q}^{j}}},
\end{equation}

where 

\begin{equation}
    H_{ij} := \frac{\partial^{2} L}{\partial \dot{q}^{i} \partial \dot{q}^{j}}
\end{equation}

define the components of the Hessian matrix.

We are now in a position to calculate the LV Lagrangian: we find 

\begin{equation}
\label{eq:lv_lagrangian_simplified_form}
    L_{\text{LV}} = \parentheses{\dot{q} + \gamma} \ln \parentheses{\frac{\dot{q} + \gamma}{\delta e}} - \parentheses{\beta e^{q} - \alpha q},
\end{equation}

where $e$ above is Euler's constant. Initiating the computation of the determinant of the Hessian (a $1 \times 1$ matrix here --- a scalar) mentioned in the previous paragraph yields $ \parentheses{\dot{q} + \gamma}^{-1}$.

\subsection{Mechanical Interpretation}
\label{sec:mechanics-of-lv}

Let us now remark on the features of the LV Lagrangian as it stands above. The first observation is that it remarkably assumes a kinetic-minus-potential form, although the \quotesaround{kinetic} piece is not quadratic in velocity. (It has been argued that quadratic kinetic terms in mechanics are enforced by Newton's Laws of Motion and considerations of homogeneity and isotropy of space \cite{landau-lifshitz-mechanics}.) Secondly, it is evident that, due to the presence of the logarithm and Euler's constant, the Lagrangian may be written in several equivalent ways. That we have chosen to write it in the manner that we have above will be soon elucidated.

Let us turn to the immediate analysis enabled through the Lagrange formalism of mechanics: computation of the equations of motion. As revealed by the LV Lagrangian, there is only one degree of freedom, which is denoted $q$. Thus, there is a single Euler-Lagrange equation that we will find. Effecting the differentiation according to the standard Euler-Lagrange equation --- which remains intact because the Lagrangian only carries a dependence on $q$ and $\dot{q}$ --- we find the equation of motion for $q$ to be

\begin{equation}
\label{eq:lv-lagrangian-eoms}
    \ddot{q} = \parentheses{\alpha - \beta e^{q}} \dot{q} + \gamma \parentheses{\alpha - \beta e^{q}}.
\end{equation}

With this equation of motion --- which can also be derived by taking a second time derivative of Eq. \ref{eq:lv-hamiltonian-q-dot} and substituting in Eq. \ref{eq:lv-hamiltonian-p-dot} --- we can now discuss the mechanical picture offered by using the Lagrangian approach.

\begin{figure}[t!]
    \centering
    \begin{subfigure}{0.475\textwidth}
        \centering
        \includegraphics[width=\textwidth]{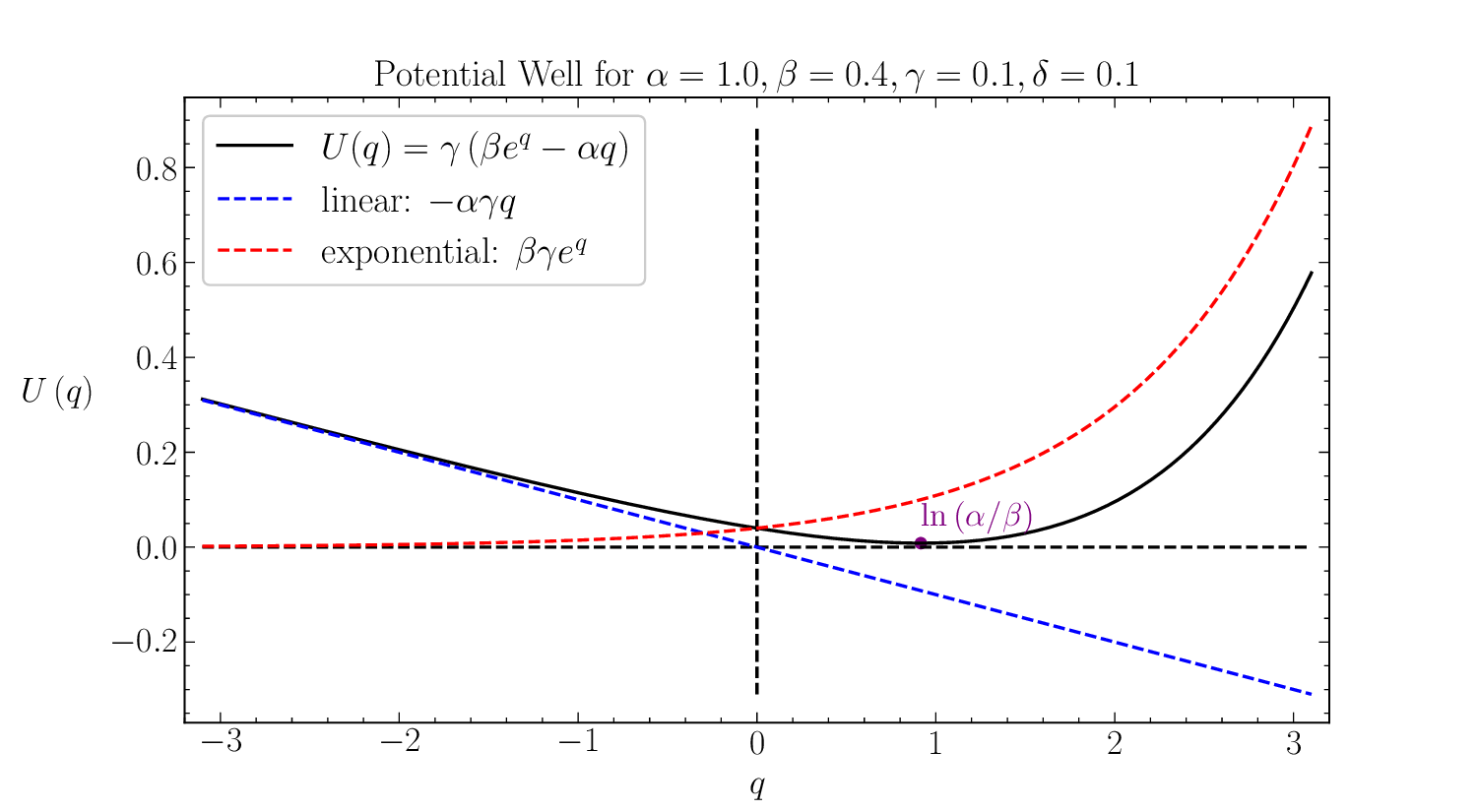}
        \caption{The contour of the potential well for the configuration of parameters $\alpha = 1.0$, $\beta = 0.4$, $\gamma = 0.1$, and $\delta = 0.1$.}
        \label{fig:lv_potential_params_alpha1_beta_04_gamma_01_delta_01}
    \end{subfigure}
    \hfill
    \begin{subfigure}{0.475\textwidth}
        \centering
        \includegraphics[width=\textwidth]{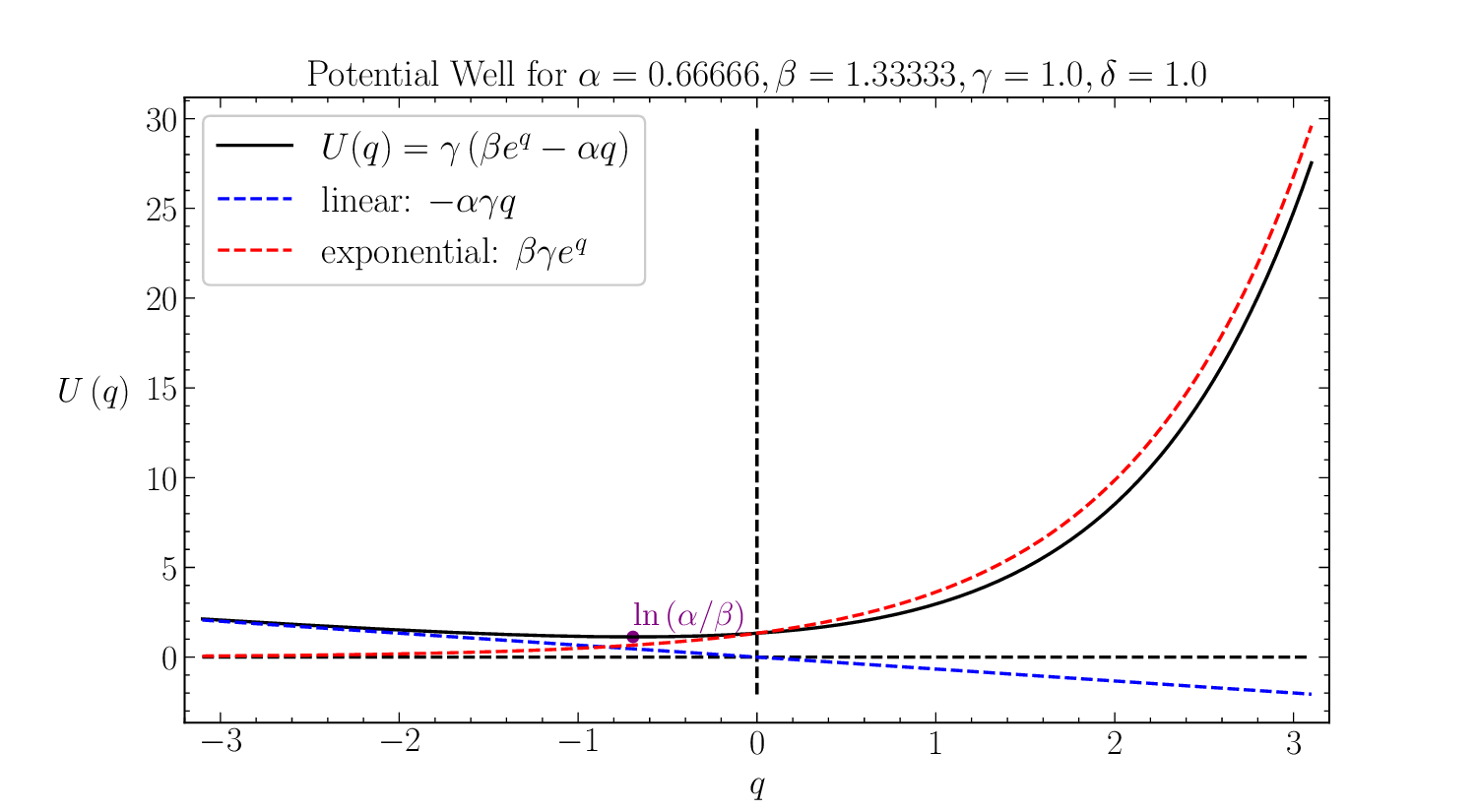} 
        \caption{The potential well for the configuration of parameters $\alpha = 0.66666 \approx \frac{2}{3}$, $\beta = 1.33333 \approx \frac{4}{3}$, $\gamma = 1.0$, and $\delta = 1.0$.}
        \label{fig:lv_potential_params_alpha06_beta_13_gamma_10_delta_10}
    \end{subfigure}
    \caption{A depiction of the potential well that the \quotesaround{predator particle} is moving in. We indicate the potential in black as the sum of two contributions: a linear piece in blue and an exponential piece in red. We also indicate the minimum of the potential using the value we derived earlier with a solid, purple dot. The coordinate zero axes are drawn in the background with gray, dashed lines. The Python code that was used to generate these figures is available open-source on a Github repository that may be found in the bibliography as Ref. \cite{lv-github-data}.}
    \label{fig:potential-wells-lv}
\end{figure}

As we have written the equation of motion above, at first glance, there appears to be a force term and a damping term. The force term corresponds to the negative gradient of a potential field of the form $U\parentheses{q} = \gamma \parentheses{\beta e^{q} - \alpha q}$; its minimum is in fact a global one, and is $q_{\text{min}} = \ln\parentheses{\alpha / \beta}$. One may expand around it to obtain the system's harmonic behavior; the harmonic expansion of $U\parentheses{q}$ yields $\tilde{U}\parentheses{q} = \alpha \gamma \parentheses{q - \ln \parentheses{\alpha / \beta}}^{2} / 2$, where we have $\tilde{U}\parentheses{q} := U''\parentheses{q_{\text{min}}}\parentheses{q - q_{\text{min}}}^{2} / 2$. One can then identify the corresponding frequencies of small oscillations around this minimum; they are $\omega = \sqrt{\alpha \gamma}$ (where we have written the equation with unit mass). Note that this Newtonian $U\parentheses{q}$ differs from the apparent potential energy that appears in the Lagrangian. We offer a depiction of this potential function in Fig. \ref{fig:potential-wells-lv}; the figures were generated using the Python programming language, using NumPy \cite{harris-2020-numpy} for numerical integration of the ODE Eq. \ref{eq:lv-lagrangian-eoms} and Matplotlib \cite{hunter-2007-matplotlib} for the visual depiction of the results of the numerical integration.

It is well known in the theory of LV that the predator equilibrium density (stable point) is wholly determined by those parameters that govern the dynamics of prey and vice versa. (One may readily verify this claim by solving for when $\dot{x} = \dot{y} = 0$ in Eqs. \ref{eq:lotka-volterra-equations}. The fixed point in the $\parentheses{x, y}$ coordinates corresponds to $\parentheses{\gamma / \delta, \alpha / \beta}$.) We have found that $q_{\text{min}}$ corresponds to the logarithm of the predator population's stable point. This logarithm is due to the coordinate transformation we performed earlier (Eq. \ref{eq:lv-y-to-q-transformation}). We have also found that the small oscillation frequency corresponds exactly to that obtained by linearization of the so-called community matrix in the study of ecology \cite{Murray2002, berlow-2004-community-matrix} around the fixed point $\alpha / \beta$. The harmonic analysis of $U\parentheses{q}$ --- as offered through this Lagrangian approach --- enabled us to provide support for the expected oscillatory behavior near the (log-transformed) LV fixed point, consistent with the linearized LV dynamics.

\begin{figure}[t!]
    \centering
    \begin{subfigure}{0.475\textwidth}
        \centering
        \includegraphics[width=\textwidth]{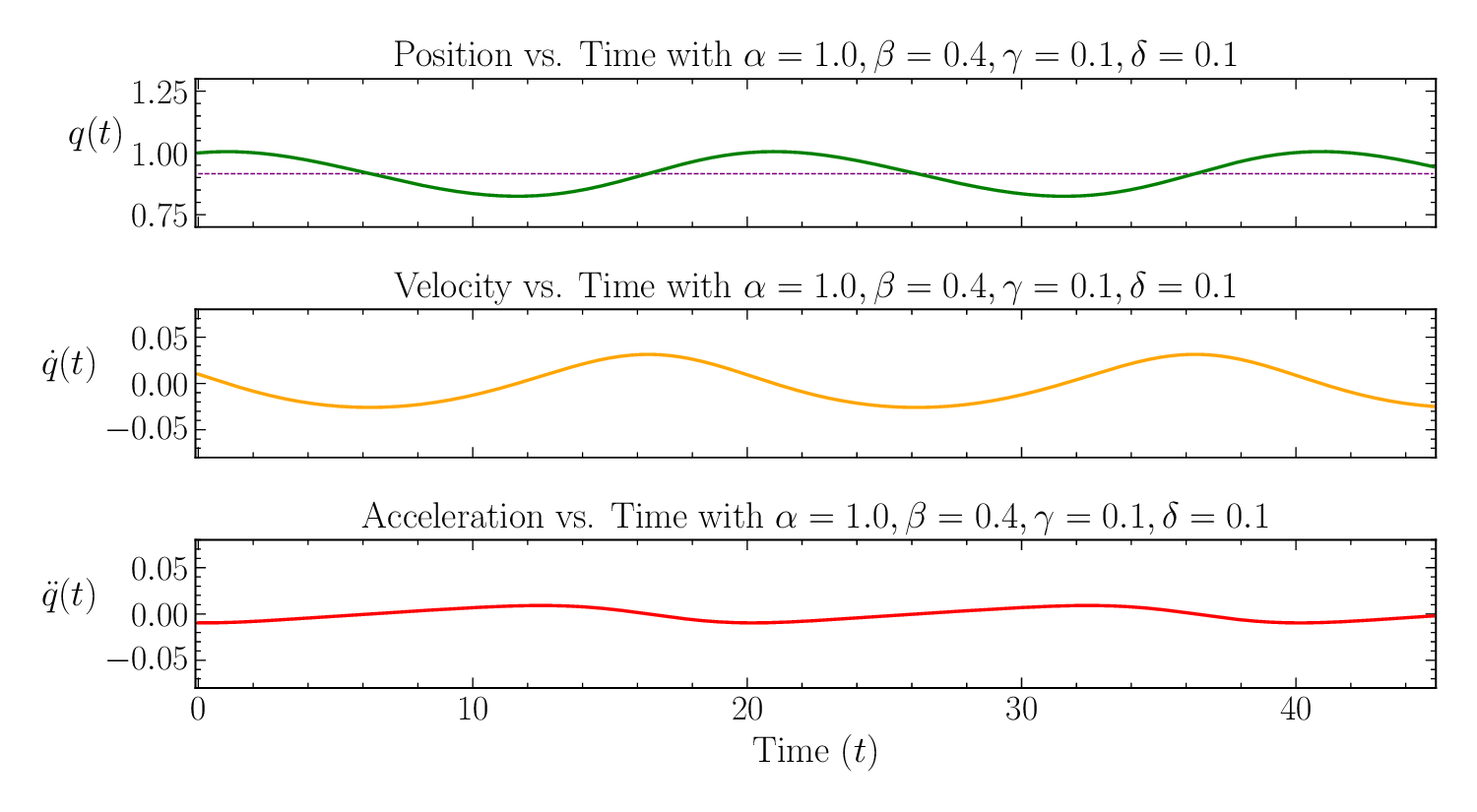}
        \caption{The kinematics of the predator particle with a parameter configuration of $\alpha = 1.0$, $\beta = 0.4$, $\gamma = 0.1$, and $\delta = 0.1$.}
        \label{fig:lv_kinematic_params_alpha1_beta_04_gamma_01_delta_01}
    \end{subfigure}
    \hfill
    \begin{subfigure}{0.475\textwidth}
        \centering
        \includegraphics[width=\textwidth]{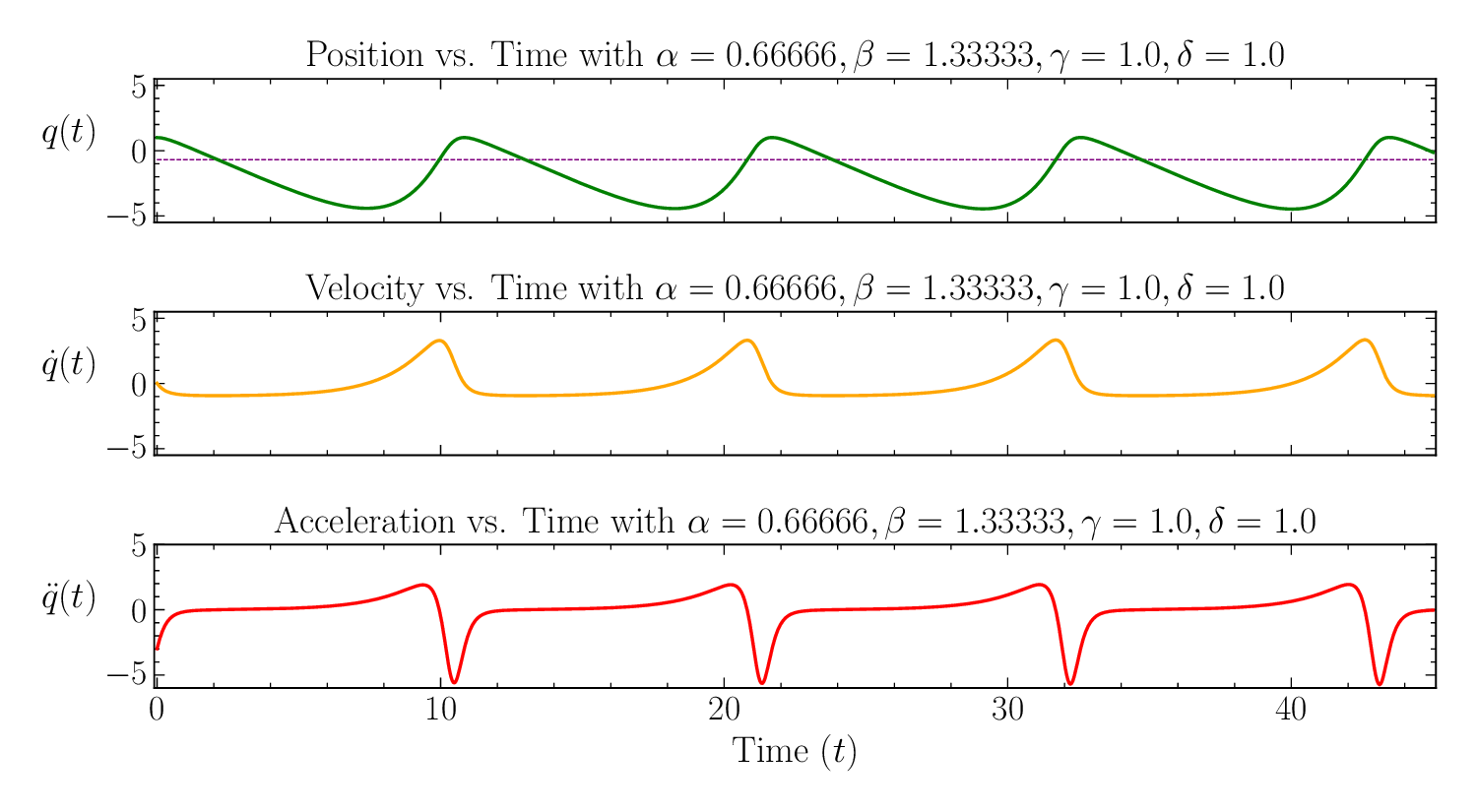}
        \caption{The kinematics of the predator particle with a parameter configuration of $\alpha = 0.66666 \approx \frac{2}{3}$, $\beta = 1.33333 \approx \frac{4}{3}$, $\gamma = 1.0$, and $\delta = 1.0$.}
        \label{fig:lv_kinematics_params_alpha06_beta_13_gamma_10_delta_10}
    \end{subfigure}
    \caption{The results of two numerical simulations of the kinematics of the \quotesaround{predator particle} according to its equation of motion (Eq. \ref{eq:lv-lagrangian-eoms}) over an interval of $\Delta T = 46$ timesteps with initial conditions $q\parentheses{0} = 1.0$ and  $\dot{q}\parentheses{0} = 0.01$. The thin, dotted purple line corresponds to the value of $q_{\text{min}}$ where the coefficient of the linear drag term flips sign.}
    \label{fig:lv-kinematics}
\end{figure}

There is an additional term in our equation of motion that acts as a linear drag on the particle. Of course, whether or not the effect is to be identified as drag or what we will hereby call \quotesaround{rev} (as in revving a motor) is determined by the overall sign of its coefficient, $\alpha - \beta e^{q}$ (which is itself a function of \quotesaround{position} of the predator density), and the sign of the particle's velocity. We can determine where the sign flip of the coefficient occurs by solving for when $\alpha - \beta e^{q} > 0$. The point at which this turnover occurs is identified as $\ln\parentheses{\alpha / \beta} > q$. That is, the turnover of this effect corresponds precisely to the minimum value here. In particular, if $q \gtrless q_{\text{min}}$, the sign of the coefficient is $\parentheses{\mp}$; at positive velocity, the particle is damped (revved). It is expected for conserved quantities to no long be conserved upon introducing damping (or revving) terms. In this case, though the Hamiltonian (Eq. \ref{eq:hamiltonian-of-lv}) is constant along the solution trajectories even with the particle experiencing these (typically) dissipative forces like linear drag. In this manner, we find that the full canonical energy remains conserved so no dissipation occurs. We offer a representation of the kinematics of the particle in Fig. \ref{fig:lv-kinematics}. As before, we used NumPy \cite{harris-2020-numpy} for numerical integration of the ODE Eq. \ref{eq:lv-lagrangian-eoms} and Matplotlib \cite{hunter-2007-matplotlib} to depict the results of the numerical integration.

\subsection{Noether's Analysis}

A major advantage of formulating a dynamical system in terms of a Lagrangian is that this framework readily admits an analysis of its symmetries. That we have constructed a Lagrangian corresponding to the original two-dimensional Lotka-Volterra system, it is an interesting avenue of inquiry to determine what the Noether procedure may yield. Since the Lagrangian we derived does not explicitly contain time, we expect that effecting the Noether procedure for time-translation symmetry should yield a conserved \quotesaround{energy.} In the standard mechanical context, time-translation symmetry in a Lagrangian yields nothing other than the Hamiltonian.

For a Lagrangian $L\parentheses{q, \dot{q}}$, the effect of infinitesimal time-translation ($t \to t + \delta t$) and demanding the Lagrangian remain the same shows that the following quantity is conserved:

\begin{equation}
\label{eq:noether-charge-time-translation-1-dimension}
    Q = \frac{\partial L}{\partial \dot{q}} \dot{q} - L.
\end{equation}

Computing the relevant ingredients using our Lagrangian shows that we identify the following quantity to be conserved:

\begin{equation}
\label{eq:noether-charge-lv-lagrangian}
    Q_{\text{LV}} =  \dot{q} - \gamma \ln \parentheses{\frac{\dot{q} + \gamma}{\delta}} + \beta e^{q} - \alpha q + \gamma.
\end{equation}

Its time-derivative is

\begin{equation}
    \dot{Q}_{\text{LV}} = \dot{q} \squarebrackets{\frac{\ddot{q}}{\dot{q} + \gamma} + \parentheses{\beta e^{q} - \alpha}}.
\end{equation}

Therefore, $\dot{Q}_{\text{LV}}$ is equal to $0$ on-shell, i.e. when its equation of motion Eq. \ref{eq:lv-lagrangian-eoms} is satisfied. Indeed, one can verify using the conjugate momentum Eq. \ref{eq:lv_conjugate_momentum} that $Q_{\text{LV}}$ is nothing but the original Lotka-Volterra Hamiltonian, $H_{\text{LV}}$. In this way, we have related the original integral of motion in the $\parentheses{x, y}$ coordinates (Eq. \ref{eq:lv-system-const-of-motion}) to correspond to the conserved charge derived with this new Lagrangian and Noether analysis.

\section{Discussion}
\label{sec:discussion}

We now provide a commentary on the peculiar features of the LV Lagrangian and its structure.

\subsection{Identifying the Energies}
\label{sec:discussion:energies}

Let us begin with the following observation: In the previous section, we identified a potential function after obtaining the equations of motion from the Lagrangian, but the $U\parentheses{q}$ we found does not correspond to the purported potential that is identifiable in the Lagrangian of Eq. \ref{eq:lv_lagrangian_simplified_form}. In other words, even though we identified the separation of a \quotesaround{kinetic piece} and a \quotesaround{potential piece} in the Lagrangian, the negative gradient of that potential piece does not account for all of the forces on the particle. More specifically, let us identify the potential piece in Eq. \ref{eq:lv_lagrangian_simplified_form} as $V\parentheses{q}:= \beta e^{q} - \alpha q$ and note that $V\parentheses{q}$ is not what we have identified as $U\parentheses{q}$ in Sec. \ref{sec:mechanics-of-lv} ($U\parentheses{q} = \gamma V\parentheses{q}$). In this regard, we have chosen to interpret the second-order equation (Eq. \ref{eq:lv-lagrangian-eoms}) through a Newtonian framework, in which the equation of motion and the forces are given rather than as derived from a Lagrangian. (Doing so enables us to better visualize the kinematics in Figs. \ref{fig:lv_kinematic_params_alpha1_beta_04_gamma_01_delta_01}, \ref{fig:lv_kinematics_params_alpha06_beta_13_gamma_10_delta_10}.)

\subsection{The $\delta$ Redundancy}

We shall now comment on a subtle feature of this construction. The LV Lagrangian includes all parameters of the original LV system, but upon application of the Euler-Lagrange equations, it is inescapable that $\delta$ drops out. It may then seem suspect that the Lagrangian formulation of the LV system fully captures the original dynamics. (One should remark, however, that $\delta$ vanishes even at the level of taking the time derivative of Eq. \ref{eq:lv-hamiltonian-q-dot}.)

In passing from the Lagrangian to the Hamiltonian, we generally expect the form of the canonical momentum to change: $p_{i} = \partial L /\partial \dot{q}^{i}$. That is, $p_{i}$ is sensitive to how a Lagrangian is written. As a brief example, if one introduces a term proportional to $q \dot{q}$ in a free Lagrangian of one degree of freedom ($L_{0} = m\dot{q}^{2}/2$), they will find that $p = m\dot{q} + q$. Of course, $q\dot{q} = d\parentheses{q^{2}/2}dt$, and thus does not affect the equations of motion. In this regard, since $p$ generally changes its definition, it is not an observable quantity. A familiar example of this idea can be found in the incorporation of electromagnetic potentials in a classical Lagrangian. The canonical momentum derived from the Lagrangian describing a charged particle in an electromagnetic field configuration, $L = m \dotproduct{\dot{\vectorarrow{x}}}{\dot{\vectorarrow{x}}}/2 + q\parentheses{\dotproduct{\dot{\vectorarrow{x}}}{\vectorarrow{A}} - \phi\parentheses{\vectorarrow{x}}}$, changes if we add a total derivative term $d \Lambda\parentheses{\vectorarrow{x}, t}/ dt$ to it, but does not affect its dynamics: the electromagnetic potentials, and consequently the canonical momentum, depend on the choice of gauge, while the electromagnetic fields and the equations of motion remain invariant.

If we require the canonical variable $p$ to retain its original identification (Eq. \ref{eq:lv-x-to-p-transformation}), and correspond to a \quotesaround{physical observable,} then the total derivative freedom cannot be used to remove the $\delta$-dependent momentum shift without changing the identification that it is a physical observable. We must decide what constitutes an observable quantity in this model and the accompanying formalism. This understanding is important because, as we alluded to in the earlier section, the Lagrangian we derived admits a myriad of different mathematical representations that all yield the same equation of motion. Given that the equation of motion omits the parameter $\delta$, it seems that these details conspire to the conclusion that information has somehow been lost in the passage from the original model to the Lagrangian. If one were to start with only the Lagrangian in Eq. \ref{eq:lv_lagrangian_simplified_form}, one may transform it according to writing the term $\dot{q} \ln \parentheses{e \delta}$ as a total time derivative of a function $f\parentheses{q, t}$, namely $f\parentheses{q, t} = q \ln\parentheses{e \delta}$; the reader will find that the canonical momentum equation (Eq. \ref{eq:lv_conjugate_momentum}) has changed accordingly, and will no longer contain $\delta$, but the equations of motion will remain untouched. Thus, in order for $p$ --- related to the prey population density --- to remain an observable quantity in our model (which it is in the original formulation of the system), we demand that Eq. \ref{eq:lv_conjugate_momentum} take the form it does without exception; this demand is enforced by not removing the otherwise redundant term $\dot{q} \ln \parentheses{e \delta}$ in Eq. \ref{eq:lv_lagrangian_simplified_form}.

On the other hand, suppose that we do remove the redundant total time derivatives from Eq. \ref{eq:lv_lagrangian_simplified_form}. The term that can be written as a total time derivative is $\parentheses{\dot{q} + \gamma} \ln \parentheses{e \delta} = d f\parentheses{q, t} / dt$, where $f\parentheses{q, t} := \parentheses{q + \gamma t} \ln\parentheses{e \delta}$. Then, we arrive at the Lagrangian

\begin{equation}
    L_{\text{LV}}' = \parentheses{\dot{q} + \gamma} \ln \parentheses{\dot{q} + \gamma} - \parentheses{\beta e^{q} - \alpha q}.
\end{equation}

We can verify that the equation of motion generated from the Lagrangian above is identical to the one derived from the previous Lagrangian (Eq. \ref{eq:lv-lagrangian-eoms}), as should be. Now, computation of the canonical momentum yields

\begin{equation}
\label{eq:canonical-momentum-reduced}
    P\parentheses{\dot{q}} = \ln\parentheses{\dot{q} + \gamma} + 1.
\end{equation}

The relation to the earlier canonical momentum Eq. \ref{eq:lv_conjugate_momentum} is $P = p + \ln \parentheses{e \delta} = p + \partial_{q} f\parentheses{q, t}$. As we spoke about earlier, given that this canonical momentum can change its definition, it is not an observable in this mechanical formulation. We can also construct the Hamiltonian for $L_{\text{LV}}'$ by following-through with the Legendre transform to arrive at

\begin{equation}
\label{eq:lv-reduced-hamiltonian}
    H_{\text{LV}}' = \frac{e^{P}}{e} - \gamma P + \beta e^{q} - \alpha q.
\end{equation}

The Hamiltonian above is of the same form as that derived earlier but with the replacement that $\delta \to e^{-1}$. Indeed, using the inverse coordinate transformations (the inverse of Eqs. \ref{eq:lv-x-to-p-transformation} and \ref{eq:lv-y-to-q-transformation}), we find the original LV equations but with $\delta \to e^{-1}$:

\begin{equation}
\label{eq:lv-reduced-equations}
    \frac{dx}{dt} = \alpha x - \beta x y, \quad
    \frac{dy}{dt} = - \gamma y + \frac{1}{e} x y.
\end{equation}

\subsection{Rotated-Coordinates Transformation}

While attempting to resolve the Delta Dropout problem above, we discovered that there exists a more symmetric and physical choice for the original logarithmic transformation. Returning to Eqs. \ref{eq:lv-x-to-p-transformation} and \ref{eq:lv-y-to-q-transformation}, one must reappraise the sentence just below the equations: \quotesaround{Note that, due to the symmetry of the transformation, the association of $x$ with $p$ and $y$ with $q$ is arbitrary.} If the association of the predator population ($y$) with the generalized coordinate $q$ is arbitrary, why did we choose to use it? Of course, what we meant to emphasize at that time was that one could easily obtain the same Lagrangian but where $q$ is now associated with $x$ instead of $y$; in that case, the Lagrangian describes the dynamics of the prey population instead of the predator population.

Instead of picking out one or the other species to serve as the generalized coordinate in the mechanical approach to this problem, let us put both on the same footing. To do so requires the standard logarithmic transformation to each dynamical variable and then effecting a rotation in this new space. Explicitly,

\begin{equation}
    p := \frac{\ln\parentheses{x} + \ln\parentheses{y}}{\sqrt{2}}, \quad q := \frac{\ln\parentheses{y} - \ln\parentheses{x}}{\sqrt{2}}.
\end{equation}

Another way to see this symmetric coordinate transformation is to canonically transform after the original transformation (Eqs. \ref{eq:lv-x-to-p-transformation}, \ref{eq:lv-y-to-q-transformation}) in the manner of rotating the variables by $\pi / 4$ in phase space. 
Note that the axis of the first transformation is decided by our definition of which term has the minus sign in the Poisson bracket. Since our mechanical system is two dimensional in phase space, any SO$\parentheses{2}$ rotation of canonical coordinates is a viable canonical transformation.

With these new coordinates, we find that the dynamical invariant Eq. \ref{eq:lv-system-const-of-motion} can be written according to 

\begin{equation}
\label{eq:hamiltonian-of-rotated-lv}
    H_{\text{LV}}'' = \delta e^{\parentheses{p - q}/\sqrt{2}} + \beta e^{\parentheses{p + q}/\sqrt{2}} - p \parentheses{\frac{\gamma + \alpha}{\sqrt{2}}} + q \parentheses{\frac{\gamma - \alpha}{\sqrt{2}}}.
\end{equation}

Not only is the above quantity a dynamical invariant, it also functions as a Hamiltonian; one can use it to derive the dynamics for the new definitions of $p$ and $q$.

\begin{align}
    \dot{q} &= \frac{\partial H_{\text{LV}}''}{\partial p} =  \frac{1}{\sqrt{2}} \parentheses{\delta e^{\parentheses{p - q}/\sqrt{2}} + \beta e^{\parentheses{p + q}/\sqrt{2}} - \gamma - \alpha}, \label{eq:lv-prime-hamiltonian-q-dot} \\
    - \dot{p} &= \frac{\partial H_{\text{LV}}''}{\partial q} =  -\frac{1}{\sqrt{2}} \parentheses{-\delta e^{\parentheses{p - q}/\sqrt{2}} + \beta e^{\parentheses{p + q}/\sqrt{2}} + \gamma - \alpha} \label{eq:lv-prime-hamiltonian-p-dot}.
\end{align}

When ready to construct the corresponding Lagrangian, one need only invert Eq. \ref{eq:lv-prime-hamiltonian-q-dot} to express $p$ as a function of $q$ and $\dot{q}$. The inversion is possible and non-singular (as can be verified by computing the Hessian, Eq. \ref{eq:hessian-definition}), and yields

\begin{equation}
    p\parentheses{q, \dot{q}} = \sqrt{2} \ln \squarebrackets{\frac{\alpha + \gamma + \sqrt{2} \dot{q}}{\delta e^{-q / \sqrt{2}} + \beta e^{q / \sqrt{2}}}}.
\end{equation}

Now that we have the map between the canonical momentum and the generalized velocity, we can proceed to construct the Lagrangian. Using Eq. \ref{eq:lagrangian-from-hamiltonian}, we find that the Lagrangian, as written in the new coordinates, is

\begin{equation}
\label{eq:lv-lagrangian-complicated-form}
    L_{\text{LV}}'' = \parentheses{\sqrt{2} \dot{q} + \alpha + \gamma} \parentheses{ \ln\squarebrackets{\frac{\sqrt{2} \dot{q} + \alpha + \gamma}{\delta e^{-q/\sqrt{2}} + \beta e^{q/\sqrt{2}}}} - 1} - q\parentheses{\frac{\gamma - \alpha}{2}}.
\end{equation}

As before, we can use the above Lagrangian to derive the equation of motion for $q$ using the standard Euler-Lagrange equation. Upon computation of the equation of motion as derived from Eq. \ref{eq:lv-lagrangian-complicated-form}, the $\delta$ parameter is preserved in the dynamics:

\begin{equation}
\label{eq:lv-lagrangian-complicated-eoms}
    \ddot{q} + \frac{1}{2}\squarebrackets{\frac{\gamma - \alpha}{\sqrt{2}} + \parentheses{\alpha + \gamma} \tanh \parentheses{\frac{q}{\sqrt{2}} + \frac{1}{2} \ln \parentheses{\frac{\beta}{\delta}}}} \parentheses{\dot{q} + \frac{\alpha + \gamma}{\sqrt{2}}} = 0,
\end{equation}

(provided $\sqrt{2} \dot{q} + \alpha + \gamma \neq 0$).

We end with a final comment on Eq. \ref{eq:lv-lagrangian-complicated-form}. It is not evident from looking at the Lagrangian that we have a separation of a \quotesaround{kinetic piece} and a \quotesaround{potential piece} that we enjoyed with the previous form (Eq. \ref{eq:lv_lagrangian_simplified_form}); the spoiling term is found by separating the denominator of the logarithm from its numerator using the standard algebraic rules of logarithms and finding $\parentheses{\sqrt{2} \dot{q} + \alpha + \gamma} \ln\parentheses{\delta e^{-q / \sqrt{2}} + \beta e^{q / \sqrt{2}}}$. Remarkably, using the dilogarithm, one can write the spoiling term as a total time derivative:

\begin{equation}
    \frac{d}{dt} \text{Li}_{2} \parentheses{-\frac{\beta}{\delta} e^{\sqrt{2} q}} = - \sqrt{2} \dot{q} \ln \parentheses{1 + \frac{\beta e^{\sqrt{2} q}}{\delta}} = - \parentheses{\sqrt{2} \dot{q} \ln \parentheses{\delta e^{-q / \sqrt{2}} + \beta e^{q / \sqrt{2}}} + \frac{d}{dt} \parentheses{\frac{q^{2}}{2} - \sqrt{2} \ln \parentheses{\delta} q}}.
\end{equation}

The total time derivative that we can remove is

\begin{equation}
    \frac{d}{dt} h\parentheses{q, t} = \frac{d}{dt} \squarebrackets{\text{Li}_{2} \parentheses{-\frac{\beta}{\delta} e^{\sqrt{2} q}} + \frac{q^{2}}{2} - \sqrt{2} \ln \parentheses{e \delta} q - \parentheses{\alpha + \gamma} t}.
\end{equation}

Therefore, one can rewrite Eq. \ref{eq:lv-lagrangian-complicated-form} without this total time derivative and still derive Eq. \ref{eq:lv-lagrangian-complicated-eoms}, which includes all of the original model parameters of Eqs. \ref{eq:lotka-volterra-equations}.

\begin{equation}
    L_\text{LV}''' = \parentheses{\sqrt{2} \dot{q} + \alpha + \gamma} \ln \parentheses{\sqrt{2} \dot{q} + \alpha + \gamma} - \parentheses{\gamma + \alpha} \ln \parentheses{\delta e^{-q / \sqrt{2}} + \beta e^{q / \sqrt{2}}} - q \parentheses{\frac{\alpha - \gamma}{2}}.
\end{equation}

Now, it is more evident that we have a kinetic-minus-potential structure, even in the Lagrangian written with the rotated coordinates. We leave as an exercise the computation of the Euler-Lagrange equations for this reduced Lagrangian to verify that its equation of motion coincides precisely with the one we derived previously, Eq. \ref{eq:lv-lagrangian-complicated-eoms} --- as it should be.

\section{Conclusion}

We have provided a Lagrangian for the most general (two-dimensional) Lotka-Volterra model of predator-prey interactions. We obtained our result by applying the extant apparatus of both Hamiltonian and Lagrangian mechanics without appealing to the results pertaining to the Inverse Problem of Lagrangian Mechanics. Computation of the equations of motion using both the Hamiltonian approach, defined by applying Hamilton's equations to Eq. \ref{eq:hamiltonian-of-lv}, and the Lagrangian approach, defined by using the Euler-Lagrange equations with the Lagrangian of Eq. \ref{eq:lv_lagrangian_simplified_form} match, demonstrating that the standard canonical Hamiltonian and Lagrangian formalisms remain mutually consistent throughout the entire computation; this fact holds true even in the rotated coordinates, in which the Hamiltonian and Lagrangian are written in Eqs. \ref{eq:hamiltonian-of-rotated-lv} and \ref{eq:lv-lagrangian-complicated-form} respectively. We have also shown that one can deduce the LV Hamiltonian by effecting the Noether procedure for infinitesimal time-translation symmetry on our derived Lagrangian.

Our work here has opened several avenues of interesting inquiry. One interesting route is an inquiry into the geometric structure of the configuration space of this particle. It is well-known that the kinetic energy associated with a pseudo-Riemannian manifold can be expressed with a homogeneous quadratic form \cite{goldstein-2011-classical, landau-lifshitz-mechanics, arnold-1989-mathematical}; on a Riemannian manifold $\parentheses{M, g}$, this form is the metric tensor $g$. It is evident that the Lagrangian that we derived cannot be written in this fashion. A more general formulation of a Lagrangian (arc length) on a manifold is a Finsler manifold, in which the metric tensor is replaced by a Minkowski norm at every tangent space to the manifold \cite{deng-2012-finsler, wu-2005-comparisontheoremsfinslergeometry}, and perhaps this more general structure is required to make sense of the form of the Lagrangian. We spoke about in Sec. \ref{sec:discussion} how the $\delta$ parameter in the Lagrangian can be removed if we did not demand that $p$ be an observable of the physical model. We also discussed how an $\pi/4$ rotation preserved this model parameter but complicated the form of the associated Hamiltonian and Lagrangian. It is compelling to examine more closely how the dynamical system changes under a general rotation parameterized by an angle $\phi$, and uncover how the $\delta$ term behaves. Another avenue of investigation is the higher-dimensional generalization of the approach we have offered here. We spoke about extant generalizations of the two-dimensional LV equations in Sec. \ref{sec:introduction}, but we commented that these approaches typically do not respect the canonicity of Lagrangian and Hamiltonian mechanics. For example, we mentioned that the formulation of $n$-dimensional LV Hamiltonians and Lagrangians in these contexts typically are not restricted by the even-dimensionality of phase space.

Along with all of the research directions just offered, there is also the opportunity to study this Lagrangian and Hamiltonian system as a canonical mechanical system. We have not offered in this Article, for example, any investigation into canonical transformations, action-angle coordinates, adiabatic invariants, its Hamilton-Jacobi formulation, and, though it is not evident at the outset what advantages it has, routes to quantization, both through canonical quantization of its Hamiltonian and the path-integral approach using its Lagrangian.

\section{Statements and Declarations}

\noindent\boldface{Acknowledgments} The authors thank Ravisankar Rajagopal, Gary Quaresima, Joshua Bautisa, Peter Arnold, and Victor Guedes for insightful discussion, and Ravisankar Rajagopal, Gary Quaresima, Joshua Bautisa for invaluable peer review.

\noindent\boldface{Data Availability} The code that was used to conduct the simulations, generate the figures, and verify the calculations in this paper is available on an open-source GitHub repository \cite{lv-github-data}. 

\noindent\boldface{Competing Interests} The authors have no relevant financial or non-financial interests to disclose.

\noindent\boldface{Research Funding}
The authors did not receive support from any organization for the submitted work.

\noindent\boldface{Ethics, Consent to Participate, and Consent to Publish} 
Not applicable.

\bibliographystyle{unsrt}

\bibliography{references}

\end{document}